\documentclass[%
reprint,
showpacs,preprintnumbers,
nofootinbib,
nobibnotes,
amsmath,amssymb,
prb,
floatfix,
longbibliography
]{revtex4-2}

\usepackage{graphicx}% Include figure files
\usepackage{dcolumn}% Align table columns on decimal point
\usepackage{bm}% bold math
\usepackage{colortbl}
\usepackage[unicode=true,colorlinks=true,citecolor=blue,urlcolor=blue]{hyperref}
\usepackage{braket}
\usepackage[normalem]{ulem}
\usepackage{enumitem}

\usepackage{color}

\graphicspath{{img/}}

\begin{document}

%\title{Classical behavior of quantum systems due to the reservoir induced coherence effects}

\title{Coherent effects in quantum transport models and their classical counterparts}

\author{N.~S.~Maslova}
\affiliation{Lomonosov Moscow State University, 119991 Moscow, Russia}
\author{V.~N.~Mantsevich}
\affiliation{Lomonosov Moscow State University, 119991 Moscow, Russia}
\author{I.~M. Sokolov}
\affiliation{Humboldt University Berlin, 12489 Berlin, Germany}
\author{P.~I. Arseyev}
\affiliation{P.N. Lebedev Physical Institute RAS, 119991 Moscow, Russia}

\begin{abstract}
We analyze the transport properties of quasiparticles locally excited at an initial time moment in several exactly solvable quantum models. It is revealed that, in the investigated quantum systems, the time-dependent probability distribution function (PDF) exhibits behavior similar to that of classical continuous-time random walk (CTRW) models, such as L\'evy walks or diffusing diffusivity. For initially excited quasiparticles interacting with the same two-dimensional phonon reservoir, the exact quantum PDF reveals a U-shaped profile due to the presence of a strong ballistic component. This ballistic component arises because the absolute value of the quasiparticle velocity is bounded from above, allowing quantum coherence effects—stemming from the interaction with the common reservoir—to play a significant role. The results obtained for the phonon-assisted hopping quantum model are compared with three alternative approaches: (i) a simple tight-binding model, (ii) a model of a heterogeneous ensemble of chains with Gaussian random hopping amplitude, and (iii) a tight-binding model with random time-dependent hopping amplitude.
It is found that the PDF of the simple tight-binding model is similar to that of the phonon-assisted hopping model. For the heterogeneous ensemble of chains with a Gaussian distribution of hopping amplitudes, the PDF transforms from bimodal to monomodal via an intermediate trimodal shape as the parameters of the Gaussian distribution are varied. The spread of the PDF is ballistic for the first and the second models mentioned above. For the model with random time-dependent classical hopping amplitude, the PDF of quasiparticle spreading exhibits universal asymptotics with logarithmic corrections and resembles the diffusing diffusivity model, characterized by diffusive spread.
\end{abstract}

\maketitle{}

\section{Introduction}

Transport properties of quasiparticles locally excited at an initial time moment are of great interest for optoelectronics, nanophotonics and quantum optics \cite{Mueller2018,Mishra2020,Fu2023,Pokryshkin2023}. Most attention was paid to the models where the coherent transport is destroyed by the interaction with the bath \cite{Causin2022,Sheehan2024}. In the present paper, we consider the situations where the transport is purely coherent, but is modulated by external global influence akin to the classical diffusive diffusivity model, or when it is phonon-assisted but the phonon reservoir at very low temperature does not lead to the decoherence on the characteristic times of a transport, giving rise to a behavior very similar to the one in a L\'evy walks. 

The quantum systems behavior can be very similar to the behavior of some classical stochastic models such as L\'evy walks or diffusing diffusivity. In L\'evy walks model a particle moves with a fixed velocity for some random time. Then a particle changes the direction of motion and moves during another random time interval with the same velocity \cite{Zaburdaev2015}. Duration of movement (with particular waiting time probability density function) and velocity value are the main characteristics of this model. Transport equations for spreading of excited quasi-particles have been derived in \cite{Klafter1987}. Despite of its simplicity L\'evy walks (LW) model exhibits various transport regimes including the ballistic one. For heavy tailed waiting time probability density function one can get the universal probability density function for particle position at fixed time moment \cite{Lamperti1958}. Probability distribution function (PDF) for L\'evy walks can be U- or W-shaped. Recently L\'evy walks models have been applied for interpretation of some experimentally measured quantum effects such as nanocrystals blinking \cite{Nirmal1996,Empedocles1996,Bischof2014,Protasenko2005} and behavior of cold atoms in optical lattices \cite{Zaburdaev2015}. The problems of QDs blinking can be mapped into classical L\'evy walk model in the following way \cite{Margolin2006,Kutner1997}. The measured averaged number of photocounts during a particular time interval can be associated with mean time-dependent particle position in L\'evy walk model. It is assumed that random switching of quantum dots between radiative and non-radiative states can occur. Such two different QD states can be matched to two velocity directions in classical L\'evy walks model. For power law waiting time PDF of QDs switching between two states the distribution function of QDs fluorescence intensity coincides with the experimentally measured one \cite{Zaburdaev2015}. However, previous analysis of this effect was based on semi-classical approach with phenomenologically introduced heavy-tailed waiting time probability density function. The microscopic origin of such PDF behavior remained unclear. 

Another classical model with non-Gaussian PDF of quasi-particles is a diffusing diffusivity model. Diffusing diffusivity is the model of quasi-particle motion when the diffusion coefficient in Fokker-Plank equation or transition rates between nearest neighboring sites in balance equations are random variables. Random variations of diffusion coefficient and transition rates can occur due to fluctuation of tunneling barrier between neighboring sites caused by fluctuating external potential. Non-Gaussian time dependent probability density function of excited quasi-particles can be observed as it was shown in \cite{Barkai2020}. Authors demonstrated that fluctuations of the number of steps performed at finite time in CTRW models lead to universal exponential decay of PDF with logarithmic corrections.

We analyze the transport properties of quasi-particles locally excited at an initial time in the tight-binding chain localized on a 2D substrate, whose phonons induce hopping between neighboring sites. We focus on the 2D phonon reservoir due to its unique properties compared to 1D and 3D systems. Two-dimensional crystals have transformed fundamental research across many disciplines. The discovery of graphene enabled the study and development of novel 2D materials \cite{Novoselov2004}. These materials allow heterostructures with diverse properties. One-atom-thick crystals now form a large family covering a broad range of properties, including metals ($NbSe_2$) \cite{Li2024}, semiconductors (transition metal dichalcogenides) \cite{Manzeli2017}, insulators (hexagonal boron nitride) \cite{Molaei2021}, and magnetic materials \cite{Zhang2021}. Their properties often differ greatly from their 3D counterparts. Combining several 2D crystals in vertical or planar heterostructures, held by van der Waals forces, yields far more combinations than traditional methods, enabling applications in nanoelectronics, spintronics, photonics, and optoelectronics \cite{Shanmugam2022}. Moreover, 2D materials exhibit striking physical phenomena, notably energy transport via optically excited quasi-particles—key for both condensed matter physics and optoelectronic applications. Rapid propagation is desirable for photodetectors and photovoltaics, while localization benefits light emission and single-photon sources. Importantly, in many materials, energy transport is mediated not by free charge carriers but by Coulomb-bound electron-hole pairs called excitons \cite{Frenkel1931,Gross1952}.

This is especially true for 2D semiconductors like mono- and few-layer transition-metal dichalcogenides (TMDs), which host tightly bound excitons due to strong quantum confinement and weak dielectric screening \cite{Xiao2017,Wang2018}. Excitons in TMDs can move freely across the material plane, and their transport has attracted considerable attention. Early studies showed diffusive propagation over hundreds of nm at ambient conditions \cite{Kumar2014}. It was quickly found that exciton propagation depends on the environment and disorder. Nonlinear regimes arise from exciton-exciton annihilation \cite{Kurilovich2022,Kurilovich2024} and, at low temperatures, from scattering with long-wavelength acoustic phonons \cite{Wietek2024}. While diffusive transport remains the most studied regime in 2D semiconductors, other scenarios—such as collective exciton behavior or interaction with collective phonon modes \cite{Fogler2014,Aguila2023,Mantsevich2025}—are possible and merit further study. This motivates continued investigation of non-diffusive exciton propagation in atomically thin semiconductors and analysis of various quasi-particle transport mechanisms.

Previously, experimental analysis of locally excited excitons spreading in an ordered polymer chains was performed \cite{Dubin2005}. It was shown that a single exciton state in an individual ordered polymer chain shows macroscopic quantum spatial coherence reaching tens of micrometres, limited by the chain length. A highly bound exciton was described by a one-dimensional energy band.  The photoluminescence intensity spatial profile is established very rapidly, on a short timescale compared with the effective lifetime of the exciton. Unfortunately, up to now there is no reasonable theoretical explanation of macroscopically coherent effects of exciton spreading along the chain with strong ballistic component. Further we demonstrate that this effect can be explained in the framework of generalized tight-binding model with global fluctuating hopping amplitude. 

The paper is organized as follows. In Sec.\ref{sec:TBM} we analze the spreading of locally excited at initial time moment quasi-particle within the simple tight-binding model. In Sec.\ref{sec:Hetero} we study the heterogeneous ensemble of chains characterized by the chain's own hopping amplitude. Sec.\ref{sec:DifDif} we consider the random time-dependent hopping amplitude uniform along the chain and Sec.\ref{sec:Reservoir} is devoted to the analysis of phonon assisted hopping between the neighboring sites of the chain in the case of uniform phonon reservoir for all pairs of sites. In Sec.\ref{sec:Concl} we conclude and discuss the
main results.

\section{The tight-binding model \label{sec:TBM}}

Let us start our discussion from the analysis of the PDF of quasi-particle's positions within the simple tight-binding model, a chain with the odd total number of states $N = 2k+1$ and with a constant hopping amplitude between the neighboring sites $T$.
Although the behavior of this PDF is discussed in the literature, see e.g. \cite{Cuevas, Linares}, we reproduce here some of the calculations since this model will serve
as a foundation for our further discussions. Note that Ref.~\cite{Cuevas} also discusses the relation of the tight-binding model with the more general quantum walk problems. 

The Hamiltonian of the chain in the tight-binding approximation reads
\begin{eqnarray}
\hat{H}=\sum_{i=1}^{N}T\hat{c}_{i}^{\dag}\hat{c}_{i+1}+h.c.,
\end{eqnarray}
where the operator $\hat{c}_{i}^{\dag}$ creates a quasi-particle at the site $i$ (in a localized state $|i \rangle$) and operator $\hat{c}_{i+1}$ annihilates the quasi-particle at the neighboring site $i+1$. 
The particle starts at the central site of the chain, with the number $n = k+1$. In what follows, the value of $T$ is taken real for simplicity.

The transformation diagonalizing the Hamiltonian depends on the boundary conditions assumed. For a chain with free ends (``open boundary conditions (OBC)'' in the terminology of Ref.\cite{Penson}), one adds two additional
cites 0 and $N+1$ at which the amplitude of the wave function vanishes identically. For a periodic chain one identifies sites 1 and $N+1$.
In the basis of localized states $|i \rangle$ the matrix $\mathbf{H}$ representing $\hat{H}$ is a Toeplitz tridiagonal matrix in the case of a closed chain, and a circulant matrix for the periodic one. 
The normalized eigenvectors of these matrices are known: These are the standing waves
\begin{equation}
|\psi_m \rangle = \sqrt{\frac{2}{N+1}}\left( \sin\left(\frac{\pi m}{N+1} \right), ... \, , \sin\left(\frac{\pi m}{N+1} N \right) \right)^{\mathrm{T}}
\label{eq:ev1}
\end{equation}
($m=1,2,... ,N$) for the closed chain, and Bloch waves
\[
 |\psi_k \rangle = \sqrt{\frac{1}{N}}\left( 1, e^{ik}, ... \, , e^{ik(N-1)} \right)^{\mathrm{T}}
\]
($k=1,2,... ,N$) for the periodic one. Although the intermediate results look slightly different for these two cases, the asymptotic behavior for $N \gg 1$ is the same. In what follows we present only
the calculations for the case described by Eq.~(\ref{eq:ev1}). It is important to note that the eigenvectors do not depend on $T$, so that the 
matrices $\mathbf{H}_T$ (operators $\hat{H}_T$) pertinent to different values of $T$ commute. The eigenvalue of the matrix $\mathbf{H}_T$ to the eigenvector $| \psi_m \rangle$ is
\begin{equation}
 E_m = 2 T \cos\left(\frac{\pi}{N+1} m\right).
 \label{eq:Em}
\end{equation}
The orthogonal matrix $\mathbf{O}$ performing the transformation consists of the eigenvectors, Eq.~(\ref{eq:ev1}), taken as the columns. 
Thus, the Hamiltonian can be diagonalized by means of the following transformation
\begin{eqnarray}
\hat{c}_{j}= \sqrt{\frac{2}{N+1}}  \sum_{m}\sin\left(m\phi j\right)\hat{\psi}_{m},
\end{eqnarray}
where $\phi=\frac{\pi}{N+1}$, c.f. \cite{Penson}, leading to a diagonal representation 
\begin{eqnarray}
\hat{H}=\sum_{m}2T\cos(\phi m)\hat{\psi}_{m}^{\dag}\hat{\psi}_{m}.
\end{eqnarray}
This form will be beneficial in Sec.~\ref{sec:Reservoir}, where it will be convenient to change to a second quantization representation.

Now we turn to our initial value problem where we start from $| \psi(0) \rangle =(0,...,0,1,0,...0)^\mathrm{T}$.
Then we have 
\[
|\psi(t) \rangle = \sum_{m=1}^N c_m | \psi_m (t) \rangle,
\]
with 
\[
c_m = \langle \psi_m | \psi(0) \rangle = \sqrt{\frac{2}{N+1}} \sin\left(\frac{\pi m (k+1)}{N+1} \right),
\]
where we restored the time dependence of the wave functions: 
\begin{eqnarray}
&& | \psi_m (t) \rangle \equiv | \psi_m \rangle  \exp \left(- i \frac{E_m}{\hbar} t\right) \nonumber \\
&& = | \psi_m \rangle  e^{- 2 i T t \cos\left(\frac{\pi m}{N+1}\right)} .
\label{eq:WF0}
\end{eqnarray}
Here we have taken $\hbar = 1$, which means that $T$ is measured in the units of frequency. The combination $2 Tt$ in the exponential will be denoted by $z$ in what follows.
In the situation with time-dependent $T(t)$, the time evolution of the wave function is given by 
$|\psi(t) \rangle = e^{- i \int_0^t \hat{H}_{T(t)} dt} |\psi(0) \rangle$, where all $\hat{H}_{T(t)}$ commute making the time-ordering in the integral superfluous.
This corresponds to the form similar to Eq.~(\ref{eq:WF0}), with the only difference that the product $z = 2Tt$ 
in the exponential is changed for the time integral $z = 2 \int_0^t T(t') dt'$, so that all following equations hold true also for this case.

Now it is convenient to use the symmetry of the situation, and to renumber the sites by the index $j = n - (k+1)$, with
$j= -k, ..., 0, ... , k$. The entries of the vector $|\psi(t) \rangle = c_m |\psi_m(t) \rangle$ now read:
\begin{eqnarray*}
&& |\psi(t) \rangle_j = \langle j | \psi(t) \rangle \\ &&=\frac{2}{N+1} \sin\left(\phi m (k+1) \right) \sin\left(\phi m (k+1 + j) \right) e^{- i z \cos \left(\phi m \right)} \\
&&= \frac{1}{N+1} \left[ \cos(\phi mj) - \cos[\phi m (2k+2+j)] \right] e^{- i z \cos \left(\phi m \right)}
\end{eqnarray*}
with $j=-k, ..., k$. Taking into account that $\phi = \frac{\pi}{N+1} = \frac{\pi}{2k+2}$ we obtain
\begin{eqnarray}
|\psi(t) \rangle_j &=& \frac{1}{N+1} \left[ \cos(\phi m j) - \cos(\phi m j + \pi m) \right] e^{- i z \cos \left(\phi m \right)} \nonumber \\
  &=& \frac{2}{N+1} \cos (\phi m j) e^{- i z \cos \left(\phi m \right)} \left\{
  \begin{array}{cc}
   1, & m \mbox{ odd} \\
   0, & m \mbox{ even} 
  \end{array} \right. . \label{eq:WF2}
\end{eqnarray}
Alternatively, one can write
\[
 |\psi(t) \rangle_j = \frac{2}{N+1} \cos (\phi m j) \sin^2 \left(\frac{\pi m}{2} \right) e^{- i z \cos \left(\phi m \right)} 
\]
where the sinusoidal term is 1 for $m$ odd, and vanishes for $m$ even. 
The corresponding probabilities to find a particle at the site $j$ are given by 
\begin{eqnarray}
 && P_j(t) = |\langle j | \psi(t) \rangle |^2 \label{eq:Expl} \\
 &&= \frac{4}{(N+1)^2} \sum_{m} \sum_{m'}  \cos (\phi m j) \cos (\phi m' j) e^{i z [\cos \left(\phi m' \right) - \cos \left(\phi m \right)]} \nonumber
\end{eqnarray}
where in the last line the summation follows over the odd values of $m$ and $m'$.

\begin{figure}[h]
\centering
\includegraphics[width=\linewidth]{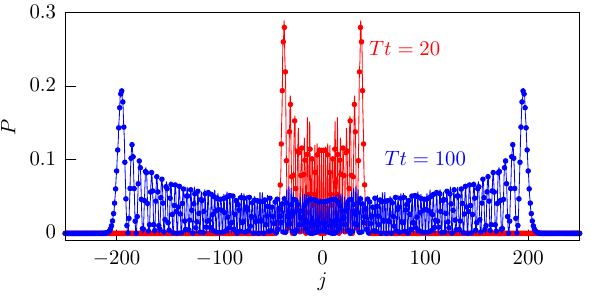}%
\caption{Quasi-particle probability density function for a tight-binding chain with regular hopping amplitude with different values of parameter $Tt$, see text for details. The dots correspond to the physically realizable
whole values of $j$, while thin solid
lines serving as a guide to the eye are obtained by evaluation of the sum, here given by Eq.~(\ref{eq:Expl}), for fractional values of $j$. The same notation is used in Figs.~\ref{fig:fig_classic_noise}, \ref{fig:fig_phonons}, and \ref{fig:fig_phonons_phase}. } 
\label{fig:fig_simple_chain}
\end{figure}

The PDF obtained by means of tight-binding model with regular amplitude, as following by explicit summation in Eq.~(\ref{eq:Expl}) with $N = 1025$ is shown in Fig.~\ref{fig:fig_simple_chain}. The time dependent PDF in this case is U-shaped and is very similar to the L\'evy walks model. Results obtained within the tight binding model differ from effective mass approximation of free quasi-particles spreading which demonstrate typical 
diffusive-like motion. 

In the tight-binding model both the energy spectrum and the  absolute value of quasi-particles' velocity are bounded from above. This is the reason for time dependent PDF resembling classical L\'evy walks model. Let us note, that the mechanism of quasi-particles spreading is different from the classical L\'evy walks because there is no real scattering as assumed in L\'evy walks model. Nevertheless, the strong similarities in PDF behavior are evident.

Although the PDF of displacements in the tight-binding model looks quite similar to the bimodal, U-shaped PDF of a classical L\'evy walk (overlayed by quantum oscillations), and the models are indeed somewhat related, this relation is not very close,
since the tight-binding model does not include any scattering. Classically, the behavior seen in the tight-binding model corresponds to a spread of particles' packet with a velocity spectrum corresponding to the group velocities 
of quantum particles. For a concentrated initial condition, the velocities are $v(\kappa) = 2 T a \sin(\kappa)$ with $a$ here being a lattice constant, and $\kappa$ homogeneously distributed on $[-\pi,\pi]$, which explains the arcsine law.

The model of 
phonon-assisted hopping between neighboring sites in presence of the zero-temperature phonon reservoir considered in Sec.~\ref{sec:Reservoir} presents, however, a much closer quantum  
analogue of a L\'evy walks model, due to the presence of real scattering of quasi-particles. 

Now, two additional notes are in place. First we note that for $N \gg 1$ the sums in Eq.~(\ref{eq:WF2}) (and also in Eq.~(\ref{eq:Expl})) can be changed for integrals. For example, in Eq.~(\ref{eq:WF2}) we get:
\[
\langle j |\psi(t) \rangle \simeq \frac{1}{\pi} \int_0^\pi d\theta \cos({\theta j} )e^{i z \cos \theta} = i^j J_j(z),
\]
see Eq.~(9.1.21) of Ref.~\cite{AbraSteg}. Therefore, the double sum in Eq.~(\ref{eq:Expl}) is represented as 
\[
|S|^2 =| i^j J_j(z)|^2 = J^2_j (z).
\]
Noting that $J_j(-z) = (-1)^j J_j(z)$ we see that this expression is invariant w.r.t. the sign of $T$. The overall normalization is correct by virtue of the addition theorem
\[
\sum_{n=-\infty}^\infty J_n^2(z) = 1,
\]
see Eq.~(9.1.76) of \cite{AbraSteg}.
We thus have
\begin{equation}
P_j(t) = J^2_j (2Tt)
\label{eq:Bess}
\end{equation}
for the case of the constant $T$. We will consider the argument $z = 2 T t$ large, in which case the distribution $P_j$ is broad. As shown in Fig. 1, this distribution is symmetric with respect 
to the origin, has a width of the order of $4 T t$ with peaks around $\pm 2 T t$, and rapidly decays outside of the corresponding interval. 

We can consider the spread $W$ of the corresponding distribution, i.e. the variance
\begin{equation}
 W^2(z) = \sum_{j=-\infty}^\infty j^2 P_j(z) = 2 \sum_{k=1}^\infty k^2 J_k^2(z) = \frac{z^2}{2}
 \label{eq:Width}
\end{equation}
(the corresponding infinite sum is given by Eq.~(5.7.12.21) of Ref.~\cite{BPM2}). Taking into account that $z=2Tt$, we see that the spread of the distribution is ballistic.

Let us discuss an alternative derivation of the same result, for a more general case, which will be used in the in the following sections.

We note that after passing to the integral in Eq.~(\ref{eq:Expl}) we get
\begin{equation}
	P_j(t) \simeq \frac{1}{\pi^2} \int_0^\pi \int_0^\pi d\theta d\theta'  \cos (\theta j) \cos (\theta' j) e^{i z [\cos \left(\theta \right) - \cos \left(\theta' \right)]}.
	\label{eq:IntRep}
\end{equation}
In Secs. \ref{sec:Hetero}, \ref{sec:DifDif} and \ref{sec:Reservoir} we will encounter more general but similar forms
\begin{eqnarray*}
	P_j && =  \frac{1}{\pi^2} \int_0^\pi \int_0^\pi d\theta d\theta' \cos(\theta j) \cos(\theta' j) \\
	&& \times e^{iA[\cos(q \theta)-\cos(q \theta')]- B[\cos(\theta)-\cos(\theta')]^2},
\end{eqnarray*}
with $q=1,2$ (the value $q=2$ will be needed in Sec. V, the parameters $A$ and $B$ have different meaning and take different values in different sections), in the present section we have $q=1$, $A=z$ and $B=0$.
Now we extend the domain of integration to the full period of the cosines,
\begin{eqnarray*}
	P_j && =  \frac{1}{4 \pi^2} \int_{-\pi}^\pi \int_{-\pi}^\pi d\theta d\theta' \cos(\theta j) \cos(\theta' j) \\
	&& \times e^{iA[\cos(q \theta)-\cos(q \theta')]- B[\cos(\theta)-\cos(\theta')]^2},
\end{eqnarray*}
and calculate
\begin{eqnarray}
	W^2 &=&\sum_{j=-\infty}^\infty j^2 P_j \nonumber  \\
	&=& \frac{1}{4 \pi^2} \int_{-\pi}^\pi \int_{-\pi}^\pi  d\theta d\theta' 2 \left[\sum_{j=0}^\infty j^2 \cos(\theta j) \cos(\theta' j)  \right] \nonumber \\
	&&  \times e^{iA[\cos(q \theta)-\cos(q \theta')]- B[\cos(\theta)-\cos(\theta')]^2} \label{eq:Width1}
\end{eqnarray}
(note that in the middle line we changed the lower limit of summation and doubled the sum). We note that the function $f(\theta,\theta') = e^{iA[\cos(q \theta)-\cos(q \theta')]- B[\cos(\theta)-\cos(\theta')]^2}$ is an even 
function of $\theta$ for any $\theta'$, so that the integrals $\int_{-\pi}^\pi f(\theta,\theta') \sin(\theta j) d\theta$ vanish, and that the sum in Eq.~(\ref{eq:Width1}) can be augmented by sinusoidal terms
$\sin(\theta j) \sin(\theta'j)$ whose contribution vanishes under integration. Thus we may write $\sum_{j=0}^\infty j^2 \cos(\theta j) \cos(\theta' j) + \sum_{j=1}^\infty j^2 \sin(\theta j) \sin(\theta' j)$ in the square brackets in
the middle line of Eq.~(\ref{eq:Width1}).

Now we note that the cosine and sine functions build a complete orthonormal system of functions (a Fourier basis) on $[-\pi, \pi]$. The properly normalized functions are 
\begin{eqnarray*}
	\phi_0(\theta) &=& \frac{1}{\sqrt{2\pi}}, \\
	\phi_j(\theta) &=& \frac{1}{\sqrt{\pi}} \cos(\theta j), \qquad j \geq 1 \\
	\widetilde{\phi}_j(\theta)&=& \frac{1}{\sqrt{\pi}} \sin(\theta j) .
\end{eqnarray*}
The sum $\sum_{j=0}^\infty |\phi_j(\theta) \rangle \langle \phi_j(\theta') | + \sum_{j=1}^\infty |\widetilde{\phi}_j(\theta) \rangle \langle \widetilde{\phi}_j(\theta') | = \delta(\theta-\theta')$ 
represents the unit operator, and our augmented sum is therefore given by
\begin{eqnarray*}
	&& \sum_{j=0}^\infty j^2 [\cos(\theta j) \cos(\theta' j) + \sin(\theta j) \sin(\theta' j) ]  \\
	&& = \frac{d^2}{d \theta^2} \pi \left[\sum_{j=0}^\infty |\phi_j(\theta) \rangle \langle \phi_j(\theta') | + \sum_{j=1}^\infty |\widetilde{\phi}_j(\theta) \rangle \langle \widetilde{\phi}_j(\theta') |\right]\\ 
	&& = - \pi \frac{d^2}{d \theta^2} \delta(\theta-\theta').  
\end{eqnarray*}
Now we turn to our integral in Eq.~(\ref{eq:Width1}) and use the property of the derivatives of a delta-function: 
\begin{eqnarray*}
	W^2 &=& - \frac{1}{2 \pi}  \int_{-\pi}^\pi \int_{-\pi}^\pi  d\theta d\theta' \frac{d^2}{d \theta^2} \exp \left\{iA[\cos(q \theta)-\cos(q \theta')] \right. \\
	&& - \left. B[\cos(\theta)-\cos(\theta')]^2\right\} \delta(\theta-\theta').
\end{eqnarray*}
The second derivative involved is
\begin{eqnarray*}
	&& \frac{d^2}{d \theta^2} \exp\{iA[\cos(q \theta)-\cos(q \theta')]- B[\cos(\theta)-\cos(\theta')]^2\} = \\
	&& \qquad \exp\{... \} \left\{-i A q^2 \cos(q\theta) + 2 B \cos \theta (\cos \theta - \cos \theta')  \right.  \\
	&& \qquad  \left.  - 2 B \sin^2 \theta + [2 B (\cos \theta - \cos \theta') \sin \theta - iA \sin(q \theta)]^2\right\}. 
\end{eqnarray*}
Performing the integration over $\theta'$ (i.e., simply setting $\theta' = \theta$) we get a single integral
\begin{eqnarray}
	W^2 &=& - \frac{1}{2 \pi}  \int_0^{2 \pi} d\theta \left\{-i A q^2 \cos(q\theta) - 2 B \sin^2 \theta \right. \nonumber  \\
	&& \left. + [iA \sin(q \theta)]^2\right\} d \theta = \frac{A^2 q^2}{2} + B. \label{eq:Width2}
\end{eqnarray}
In the present discussion we have $q=1$, $A=z$ and $B=0$, in Section III  $q=1$, $A = \mu$ and $B = \sigma^2/2$. In Section IV one has $q=1$, $A=0$ and $B = 2Dt$. In Section V we have $q=2$, $A = \alpha \omega_c t$, and 
$B=\alpha \ln(\omega_c t)$ for 2D phonon reservoir, $B=\beta\omega_{c}t$ for 1D phonon reservoir and $B=\gamma$ for 3D phonon reservoir. The parameters are the following $\alpha=\frac{M^{2}}{s^{2}q_{0}^{2}}$, $\beta=\frac{M^{2}}{sq_{0}\omega_{c}}$ and $\gamma=\frac{M^{2}\omega_{c}}{s^3q_{0}^{3}}$, where $s$ is the speed of sound, $M$ is the matrix element of quasiparticle-phonon and $\omega_c$ is the edge of phonon frequency. The detailed calculation of these coefficients can be found in \cite{Maslova2023}.

We also note that the integral, Eq.~(\ref{eq:IntRep}) can be evaluated by a saddle-point approximation provided $z$ is large, giving
\begin{equation}
 P_j(z) \approx P(x|z) = \frac{1}{\pi \sqrt{z^2-x^2}},
 \label{eq:ArcSin}
\end{equation}
for $-z < x < z$, with $x$ being a continuous variable interpolating between the sites $j$. This is a classical symmetric arcsine distribution. The result is thus the same as
discussed in Refs. \cite{Cuevas, Linares} which assume, however, a circulant model. 

\section{Heterogeneous ensemble of chains \label{sec:Hetero}}

\begin{figure}[h]
\centering
\includegraphics[width=\linewidth]{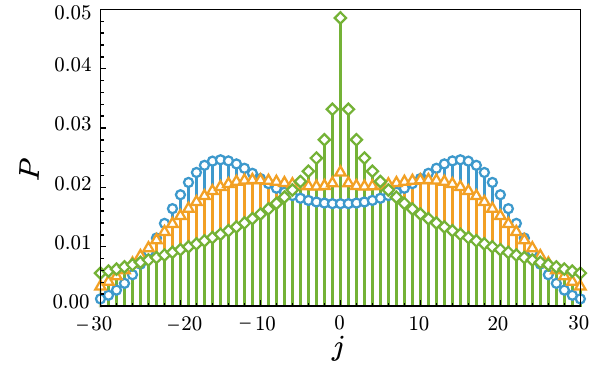}%
\caption{The probability to find a particles at a site $j$ of a chain for $\mu=20$ and different values of $\sigma$. One readily infers the transition from a bimodal shape for $\sigma$ small, $\sigma = 5$ (circles) to 
a monomodal shape for sigma large ($\sigma = 20$, diamonds). The intermediate domain (here shown for $\sigma = 7.5$, triangles) corresponds to trimodal shapes. \label{fig:Bess}}
\label{fig:hetero}
\end{figure}

As an intermediate step, let us consider a heterogeneous ensemble of chains, characterised by the chain's own hopping amplitude $T$ each. We will assume these amplitudes to follow a 
Gaussian distribution 
\[
p_T(T) = \frac{1}{\sqrt{2\pi} \sigma_T} \exp \left[-\frac{(T-\mu_T)^2}{\sigma_T^2} \right]. 
\]
This distribution of $T$ correspond to the Gaussian distribution $p(z)$ of $z$ with the mean $\mu = 2 \mu_T t$ and the variance $\sigma^2 = 4 \sigma_T^2 t^2$. Let us fix some $t$, and consider the 
probability to find a particle at a site $j$ in such an ensemble of chains. We get
\begin{equation}
 \overline{P}_j = \int_{-\infty}^\infty J^2_j (z) p(z) dz.
 \label{eq:DirInt}
\end{equation}
In Fig.~\ref{fig:Bess} we show the result for a fixed value of $\mu = 20$ and different values of $\sigma = 5, 7.5$ and 20. 

We now can consider the spread of the distribution. Averaging the result of Eq.~(\ref{eq:Width}) over the distribution of $z$ or directly applying Eq.~(\ref{eq:Width2}) we get:
\[
 W^2 = \mu^2/2 + \sigma^2/4,
\]
so that $W(t)$ grows ballistically with time, independently on whether the distribution is mono-, bi- or trimodal. 

We can get a continuous scaling approximation for the behavior shown in Fig.~\ref{fig:Bess} using Eq.~(\ref{eq:ArcSin}) which works reasonably well at $x$ of the order of $z$ or larger for $\sigma$ sufficiently large,
and an alternative approximation working for moderate $x$, but failing for $x \sim \sigma$ and larger.

In the first case, we start from the approximation
\[
 P(x) = \int_{-\infty}^\infty  \frac{\Theta(x^2-z^2)}{\pi \sqrt{z^2-x^2}} p(z) dz
\]
using Eq.~(\ref{eq:ArcSin}) instead of the square of the Bessel function, separate the integral into two, from $x$ to infinity, and from minus infinity to $-x$, and 
change the variable of integration from $z$ to $-z$ in the second integral, and write
\begin{eqnarray}
  \overline{P}(x) \approx  && \frac{1}{\pi} \int_x^\infty \frac{1}{\sqrt{z^2-x^2}} p(z)  dz \nonumber \\
  && + \frac{1}{\pi} \int_x^\infty \frac{1}{\sqrt{z^2-x^2}} p(- z)  dz, \label{eq:Pbar}
\end{eqnarray}
with two integrals of a similar form: 
\[
I_{\pm}(x) =\frac{1}{\sqrt{2} \pi^{\frac{3}{2}}} e^{-\frac{\mu^2}{2 \sigma^2}}\int_x^\infty \frac{1}{\sqrt{z^2-x^2}}\exp \left(- \frac{z^2}{2 \sigma^2} \pm \frac{\mu}{\sigma^2} z \right) dz.
\]
Now, we can rescale the variables, passing to dimensionless $\xi = x/\sigma$ and $\zeta = z/\sigma$, and introduce the coefficient of variation $q = \mu/\sigma$.
We now have 
\[
 I_\pm(x) = \frac{1}{\sqrt{2} \pi^{\frac{3}{2}}} e^{-\frac{q^2}{2}} \int_\xi^\infty \frac{1}{\sqrt{\zeta^2-\xi^2}}\exp \left(- \frac{\zeta^2}{2} \pm q \zeta \right) d \zeta.
\]
One can consider different approximations for these integrals, but already looking at the form above one sees that the PDF of $x$ depends only on the coefficient of variation $q$
of the distribution. The approximation always has a weak divergence at $x=0$ (instead of a cusp of a non-asymptotic form), which is due to insufficiency of the saddle-point approximation 
for $z$ small, but plotting the approximation, one can readily infer that it reasonably describes the transition between the trimodal and monomodal shape at $q \sim 1$ (essentially around
$q = 2.4$), so that one can safely assume that the cases with $q <1$ belong to the monomodal situation.

Note that deep in the monomodal domain, for $q \ll 1$, the integral for $ \overline{P}(\xi)$, Eq.~(\ref{eq:Pbar}) can be easily evaluated. In our dimensionless variables we have:
\begin{eqnarray}
 \overline{P}(\xi) &=& \frac{\sqrt{2}}{\pi^{\frac{3}{2}}} \int_\xi^\infty \frac{1}{\sqrt{\zeta^2-\xi^2}}\exp \left(- \frac{\zeta^2}{2}\right) d \zeta \nonumber \\
 &=& \frac{1}{\sqrt{2} \pi^{\frac{3}{2}}} e^{-\frac{\xi^2}{2}} K_0 \left(\frac{\xi^2}{2} \right), \label{eq:Symm}
\end{eqnarray}
which expression follows practically immediately from the integral representation of the modified Bessel function $K_0$. This expression makes evident the logarithmic singularity at zero
$ \overline{P}(\xi) \sim 2 \ln |\xi|$ for $|\xi| \ll 1$ being an artefact of the approximation (but disappearing after a whatever coarse-graining), and the behavior at the wings corresponding
to 
\begin{equation}
 \overline{P}(\xi) \sim \frac{1}{\sqrt{2} \pi} \frac{1}{|\xi|}e^{-\xi^2},
\end{equation}
where the asymptotics $K_0(z) \sim \sqrt{\frac{\pi}{2z}}e^{-z}$ is used. This, essentially Gaussian, decay at the wings is an artefact of the approximation, and the real decay is slower,
see Eq.~(\ref{eq:ExpCorr}).

In the second case we start from the representation given by Eq.~(\ref{eq:IntRep}), and average this integral expression over the distribution of $z$:
\begin{eqnarray*}
 P_j(z) = &&\frac{1}{\pi^2} \int_0^\pi \int_0^\pi d\theta d\theta'  \cos (\theta j) \cos (\theta' j) \times \\ 
 && \int dz p(z)  e^{i z [\cos \left(\theta \right) - \cos \left(\theta' \right)]} \\
 && \equiv \frac{1}{\pi^2} \int_0^\pi \int_0^\pi d\theta d\theta'  \cos (\theta j) \cos (\theta' j) \times \\ 
 && \langle \exp [i z [\cos \left(\theta \right) - \cos \left(\theta' \right)] \rangle_z, 
\end{eqnarray*}
where the mean in the last expression $f(\kappa)= \langle \exp [i z \kappa \rangle_z$  is a characteristic function of $p(z)$ taking at the Fourier variable equal to 
$\kappa = \cos \left(\theta \right) - \cos \left(\theta' \right)$
(one can also proceed with the same trick in the initial double sum representation, Eq.~\ref{eq:Expl}, as discussed above, and do the approximations at the final step).
For our Gaussian distribution $\ln[f(\kappa)] = i \mu \kappa - \frac{\sigma^2}{2} \kappa^2$. Now we can proceed with the saddle-point approximation for the last integral, 
and get for $\sigma \ll z$
\[
 P_j(z) \sim \left[z^2-j^2+ \frac{\sigma^2}{z^2}j^4 \right]^{-1/2},
\]
so that the PDF in the center resembles the classical arcsine distribution. This approximation shows, that the maxima of the distribution shift to 
\[
 j_{max}=z\left(1-\frac{1}{2}\frac{\sigma^{2}}{z^{2}}\right)^{1/2},
\]
and fall together for $\sigma^2 \approx 2 z^2$, confirming our previous discussion. Note that this saddle-point approximation fails to reproduce the  the monomodal regime, and the tails of the distribution
which appear as sharp cut-offs.

To elucidate the nature of the tails, we return to the integral form, Eq~(\ref{eq:DirInt}), which, for the symmetric case reads
\[
 \overline{P}_j = \frac{1}{\sqrt{2 \pi} \sigma} \int_{-\infty}^\infty J^2_j (z) e^{- \frac{z^2}{2 \sigma^2}} dz,
\]
and use the asymptotic expression of $J_j(z)$ for large positive indices $j$, $J_j(z) \sim \frac{1}{\sqrt{2 \pi j}} \left(\frac{e z}{2 j} \right)^j$, Eq.~(9.3.1) of \cite{AbraSteg}:
\begin{eqnarray*}
  \overline{P}_j &\sim&  \frac{1}{2 \pi j} \left(\frac{e}{2 j} \right)^{2j}  \frac{1}{\sqrt{2 \pi} \sigma} \int_{-\infty}^\infty z^{2j} e^{- \frac{z^2}{2 \sigma^2}} dz \\
  &=&  \frac{1}{2 \pi j} \left(\frac{e}{2 j} \right)^{2j} \frac{(2j)!}{2^j j!} \sigma^{2j}.
\end{eqnarray*}
Now we apply the Stirling formula for factorials, and get
\[
   \overline{P}_j \sim \frac{1}{2 \pi} \exp\left[- j \left(\ln j - \ln \frac{2 e}{\sigma^2} \right) - \ln j\right]
\]
with the super-exponential (but sub-Gaussian) asymptotics
\begin{equation}
  \overline{P}_j \sim \frac{1}{2 \pi} e^{- j \ln j},
  \label{eq:ExpCorr}
\end{equation}
which is essentially independent of $\sigma$. We will compare this approximation with the evaluation performed in the next section.

The aim of the present Section was to consider a model of intermediate complexity, and to present a general discussion of the types of the behavior 
possible in ensembles of tight-binding chains. We also discuss the domains of applicability and the accuracy of several approximations which will be also used in the following discussions. 
In what follows we consider two particular situations, a chain with common random time-dependent hopping amplitude,
a situation pertinent to a monomodal regime and very close in spirit to classical diffusing diffusivity models, \cite{Chubyns,ChechSok}, and a chain 
with phonon-assisted hopping in a contact with a two-dimensional phonon reservoir (with a bimodal PDF), being a close quantum relative of a L\'evy walk model,
and having interesting physics behind it, also without the connection to the L\'evy walks. The first model corresponds to a random classical time-dependent $T$,
and the second one can be considered as a special case of a situation where $T$ is operator-valued. We would like to note, that intermediate regime shown in Fig.\ref{fig:hetero} is one of the possible explanation of experimental results for exciton spreading in polymer chain obtained \cite{Dubin2005}.

\section{Classical random hopping amplitude \label{sec:DifDif}}
Let us now consider the quasi-particle hopping with random classical amplitude $\xi(t)$ assumed uniform along the chain, i.e. the same for all connected pairs of sites. 
Uniform random classical amplitude can appear due to the fluctuations of tunneling barrier between neighbouring sites caused by fluctuating external potential. The Hamiltonian of the system has the following form 
\begin{eqnarray}
\hat{H}=\sum_{j'=1}^{N}\hat{c}_{j'}^{\dag}\hat{c}_{j'+1}\xi(t)+h.c.,
\end{eqnarray}
where $j'=1\ldots N$ and $N=2k+1$. Here we will concentrate on a monomodal situation, and consider random classical amplitude $\xi(t)$ to be Gaussian with zero mean and 
with with correlation function $\langle\xi(t)\xi(t')\rangle=K(t-t')$. Therefore, in this section we consider specifically the monomodal regime, and provide a different approximation, specially suited for this case.

The Hamiltonian can be diagonalized similarly to the previous section. Introducing
\begin{eqnarray}
\hat{c}_{j}= \sqrt{\frac{2}{N+1}} \sum_{m}\sin\left(m\phi j\right)\hat{\psi}_{m},
\end{eqnarray}
one can get
\begin{eqnarray}
\hat{H}=\sum_{m}2\xi(t)\cos(\phi m)\hat{\psi}_{m}^{\dag}\hat{\psi}_{m}.
\end{eqnarray}
The initial state corresponds to the situation when the quasi-particle is located at site $j'=k+1$. After indexes changing with $j'=k+1+j$ one can get
\begin{eqnarray}
\psi(0)=\delta(j'-(k+1))=\delta(j)=\hat{c}_{0}^{\dag}|g\rangle.
\end{eqnarray}
The initial state with quasi-particle localized at site $j$ reads
\begin{eqnarray}
\psi_{j}=c_{j}^{\dag}|g\rangle,
\label{eqn:39}
\end{eqnarray}
where $|g\rangle$ is the ground state.
The time dependent quasi-particle spatial probability density function can be expressed as it was done in the previous section
\begin{eqnarray}
P_{j}(t)=|\langle j|\psi(t)\rangle|^{2}.
\end{eqnarray}
The explicit form of the spatial probability density function is
\begin{eqnarray}
P_{j}(t) && =\sum_{mm'}\cos(\phi mj)\cos(\phi m'j)\sin^{2}\left(\frac{\pi m}{2}\right)\sin^{2}\left(\frac{\pi m'}{2}\right) \nonumber \\
&& \times\left\langle \textrm{exp} \left[i2\int_{0}^{t}\xi(t')dt'(\cos(\phi m)-\cos(\phi m')) \right]\right\rangle_{\xi} \label{eqn:41}
\end{eqnarray}
(where we return to our note about the commutativity of the time-dependent Hamiltonians which makes the time-ordering in the last expression  superfluous). 
For $\xi(t)$ being a Gaussian random variable,
\begin{eqnarray*}
P_{j}(t)&&=\sum_{mm'}\cos(\phi mj)\cos(\phi m'j)\\
&&\times \textrm{exp}[-a(\cos(\phi m)-\cos(\phi m'))^{2}] \\
\end{eqnarray*}
(with the summation over odd $m,\, m'$) where 
\begin{eqnarray}
a=2\int_{0}^{t}\int_{0}^{t}K(t-t')dtdt'.
\label{eqn:43}
\end{eqnarray}
Changing $\sum_{mm'}\rightarrow\int dxdy$ in the limit $N\gg1$ one gets
\begin{eqnarray}
P_{j}(t)&&= \frac{1}{\pi^2} \int_{0}^{\pi}\int_{0}^{\pi} dxdy\cos(xj)\cos(yj) \nonumber\\
&&\times \textrm{exp}[-a(\cos(x)-\cos(y))^{2}]. \label{eqn:a}
\end{eqnarray}
Performing substitution $u=x+y$ and $v=x-y$ one can obtain
\begin{eqnarray}
P_{j}(t)&&=\frac{1}{4}\int_{0}^{2\pi}du\int_{-\pi}^{\pi}dv\textrm{exp}[-4a\sin^{2}(u/2)\sin^{2}(v/2)]\nonumber\\&&\cdot(\cos(uj)+\cos(vj))\nonumber\\
&&=\frac{\pi}{2}\int_{0}^{2\pi}du \textrm{e}^{-a(1-\cos(u))}I_{j}(a(1-\cos(u))).
\end{eqnarray}
Substituting $1-\cos(u)=x$ one can get
\begin{eqnarray}
P_{j}(t)&&=\pi\int_{0}^{2}I_{j}(ax)\textrm{e}^{-ax}\frac{1}{\sqrt{x(2-x)}}dx.
\end{eqnarray}
After integration, see Eq.~2.15.4.1 of \cite{BPM2},  one obtains the following result: 
\begin{eqnarray}
P_{j}(t)&&=\pi\textrm{e}^{j\ln(a)}\frac{\Gamma(j+1/2)\Gamma(1/2)}{\Gamma^{2}(j+1)}\nonumber\\&&\times _{2}F_{2}(j+1/2,j+1/2,2j+1,j+1,4a).
\label{eqn:47}
\end{eqnarray}
Asymptotics for $j\gg1$ reads
\begin{equation}
P_{j}(t)\simeq\textrm{exp}\left[-j\left[\ln\left(\frac{j+2a+2w}{-j+2a+2w}\right)+1\right]+2w\right],
\end{equation}
where $w=\sqrt{j^{2}+a^{2}}$. Finally, for $j\ll a$
\begin{eqnarray}
P_{j}(t)\simeq\textrm{exp}\left[-j\left[\ln\left(1+\frac{j}{2a}\right)+1\right]\right]
\end{eqnarray}
and for $j\gg a$
\begin{eqnarray}
P_{j}(t)\simeq\textrm{exp}\left[-j\ln\left(\frac{j}{a}\right)\right], \label{eq:Superexp}
\end{eqnarray}
cf. Eq.~(\ref{eq:ExpCorr}). The probability density function for large displacements (large values of$j$) in this quantum model has universal exponential asymptotics with logarithmic corrections similar to \cite{Barkai2020}.

\begin{figure}[h]
\centering
\includegraphics[width=\linewidth]{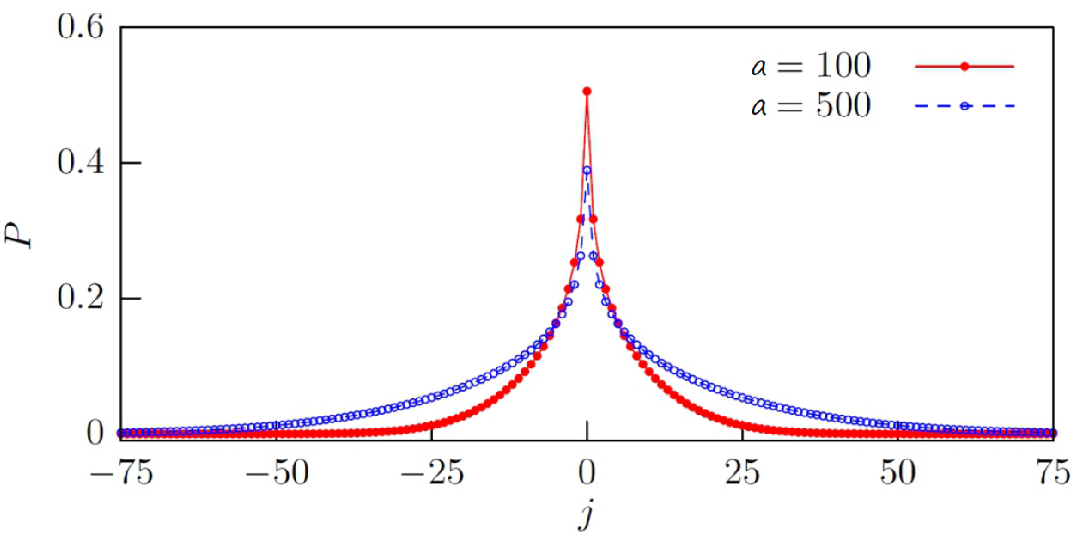}%
\caption{Quasi-particle probability density function for classical random hopping amplitude uniform for all pairs of sites for different values of parameter $a$ given by Eq.(\ref{eqn:43}).  
} 
\label{fig:fig_classic_noise}
\end{figure}

In the case of classical random hopping amplitude uniform for all pairs of sites, the time-dependent quasi-particle probability density function differs strongly from simple diffusive approximation even for delta-correlated hopping amplitude,
see below. This occurs due to the presence of correlations between hopping processes between different pairs of neighboring sites.
Note that for an integrable correlation function $K(t)$, i.e., for the classical noise having a finite correlation time, the time dependence of $a$ is linear: $a(t) \sim 2 D t$ where the ``diffusion coefficient'' $D$ is connected 
with the intensity and correlation time of the noise. The width $W^2(t)$ of the distribution at time $t$ is given by Eq.~(\ref{eq:Width2}) with $A=0$ and $B = a(t)$, and grows diffusively at all times, as $2 D t$.
This behavior is very similar to the one in a classical diffusing diffusivity models \cite{Chubyns,ChechSok} although our model is essentially a  
``diffusing velocity'' and not a diffusing diffusivity one.

Let us compare our model with a purely coherent quantum transport with its (semi)classical counterparts, in which the coherence is destroyed. In these models, the time dependence of probabilities $P_i(t)$ to find a particle
on a site $i$ of a chain is given by a master equation
\begin{equation}
 \dot{P}_{i}(t)=\sum_{i'=i \pm 1} [ w_{i,i'}P_{i'}(t) - w_{i,i'}P_{i}(t)]
\end{equation}
where the transition rates $w_{i,i'}$ are given by the Golden Rule and proportional to $|T|^2$ \cite{Zwanzig}. 
In the continuum limit in space (and at longer times) the solution to the master equation can be approximated by the one of the diffusion equation
\begin{equation}
  \dot{p}(x,t)= D \Delta p(x), 
\end{equation}
for a continuous PDF $p(x,t)$ with the diffusion coefficient $D=w a^2$ where $a$ here is the lattice constant, and $w=w_{i,i'}$ provided the transition rates between each pair of neighboring sites are the same, i.e. $D = c |T|^2$, 
with $c$ being a constant with dimension of $[c] = \mathrm{L^2}/ \mathrm{T}$ in our units in which the dimension of $T$ is the inverse time. For a homogeneous chain,
the transport is diffusive, and for a concentrated initial condition $p(x,0) = \delta(0)$ follows a Gaussian distribution
\[
 p(x,t) = \frac{1}{\sqrt{4 \pi D t}} \exp \left(-\frac{x^2}{2 D t} \right).
\]
This classical model can be considered both for an inhomogeneous ensemble of chains, and for global time-dependent transition rate. For an inhomogeneous ensemble of chains, we have
\[
 \overline{p}(x,t) = \int dD  \frac{1}{\sqrt{4 \pi D t}} \exp \left(-\frac{x^2}{2 D t} \right) p(D) dD,
\]
with $p(D)$ being the PDF of $D$. Assuming $T$ to have a Gaussian distribution with zero mean and variance $\sigma^2$, we get a Gamma-distribution
\[
 p(D) = \frac{1}{\sqrt{2 \pi c} \sigma} \frac{1}{\sqrt{D}} \exp\left(-\frac{D}{2 c \sigma^2} \right),
\]
and
\[
 \overline{p}(x,t) =  \frac{1}{\pi \sigma \sqrt{2 c t}} K_0\left(\frac{|x|}{\sqrt{2 c t} \sigma} \right)
\]
with $K_0(z)$ being a modified Bessel function of the second kind. The distribution has exponential tails, and resembles other 
cases of Brownian yet non-Gaussian diffusions, like the one of Ref.~\cite{Wang}, where the distribution of $D$ was assumed to be exponential.
This density scales as a whole, and the mean squared displacement $W^2 = \langle x^2 \rangle = 2 c \sigma^2 t$ characterizing the width of the distribution  grows diffusively, 
at difference with the coherent transport case discussed in Sec.~\ref{sec:Hetero}. 

Turning to the case when $T$ is globally time-dependent, like in Sec.~\ref{sec:DifDif}, we arrive exactly at the minimal diffusing diffusivity model of Ref.~\cite{ChechSok} if we assume that
the time dependence of $T(t)$ follows an Ornstein-Uhlenbeck process. In this case, the PDF in the classical model has a cusp and exponential tails at short times
but tends to a Gaussian at long ones. The coherent transport model in this case shows a similar diffusive growth of the width of the distribution, but no transition to a Gaussian.

\section{Phonon-assisted hopping \label{sec:Reservoir}}

We now analyze another exactly solvable model: a chain with phonon-assisted hopping of quasi-particles between neighboring sites. As discussed in the Introduction, will consider two-dimension phonon reservoir. 
The Hamiltonian of the system has the form
\begin{eqnarray}
\hat{H}=\sum_{j=1}^{N}\hat{c}_{j+1}^{\dag}\hat{c}_{j}\hat{\eta}+\hat{H}_{ph}+h.c.
\end{eqnarray}
with $\hat{c}_{j}^{\dag}$ being the creation operator and $\hat{c}_{j}$ being the annihilation operator of quasi-particle on the site $j$ of the
chain and operator $\hat{\eta}$ has the following form: 
\begin{eqnarray}
\hat{\eta}=\sum_{q}M_{q}(\hat{b}_{q}^{\dag}+\hat{b}_{q}),
\end{eqnarray}
where operator $\hat{b}_{q}^{\dag}$ describes creation and operator $\hat{b}_{q}$ describes
annihilation of phonon with $M_{q}$ being the quasi-particle-phonon coupling constant. The phonon Hamiltonian has the form
\begin{eqnarray}
\hat{H}_{ph}=\sum_{q}\omega_{q}\hat{b}_{q}^{\dag}\hat{b}_{q}
\end{eqnarray}
with $\omega_{q}$ being a phonon frequency. We consider acoustic phonons with $\omega_{q}=sq$, where $s$ is the speed of sound. The model resembles the tight-binding one, but now with the operator-valued hopping amplitude.

Diagonalization of initial Hamiltonian for quasi-particle operators can be performed in the standard way discussed above introducing the substitution
\begin{eqnarray}
\hat{c}_{j}=\sqrt{\frac{2}{N+1}}\sum_{j}\sin(m\phi j)\hat{\psi}_{m}
\label{eq:TR1}
\end{eqnarray}
with $\phi=\frac{\pi}{N+1}$. The transformed Hamiltonian reads
\begin{eqnarray}
\hat{H}_{m}=\sum_{m=1}^{N}\hat{\eta}_{m}\hat{\psi}_{m}^{\dag}\hat{\psi}_{m}+\hat{H}_{ph},
\end{eqnarray}
where
\begin{eqnarray}
\hat{\eta}_{m}=2\hat{\eta}\cos(\phi m).
\end{eqnarray}
Diagonalization leads to the renormalization of quasi-particle-phonon coupling constant $M_{q}$
\begin{eqnarray}
M_{q}\rightarrow M_{qm}=2M_{q}\cos(\phi m).
\end{eqnarray}

\begin{figure}[h]
\centering
\includegraphics[width=\linewidth]{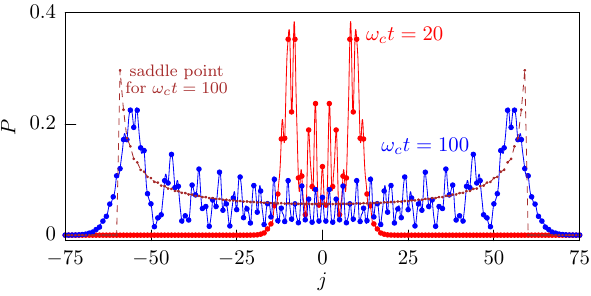}%
\caption{Quasi-particle probability density function for different values of $\omega_{c}t$ and $\delta=0$. The quasi-particles are excited at site $j=0$. The value of $\alpha$ is taken for $\alpha=0.3$. } 
\label{fig:fig_phonons}
\end{figure}

On the next step one decouples quasi-particles and phonons using the canonical transformation $\hat{H}\rightarrow\textrm{e}^{\hat{S}}\hat{H}\textrm{e}^{\hat{-S}}$ with
\begin{eqnarray}
\hat{S}=\sum_{m}\hat{\psi}_{m}^{\dag}\hat{\psi}_{m}\sum_{q}\frac{M_{qm}}{\omega_{q}}(\hat{b}_{q}^{\dag}-\hat{b}_{q}).
\end{eqnarray}
The transformed phonon operator acquires the shift of the phonon equilibrium position depending on the occupation of polaron states and can be expressed as
\begin{eqnarray}
\hat{\tilde{b}}_{q}=\hat{b}_{q}-\sum_{m}\hat{\psi}_{m}^{\dag}\hat{\psi}_{m}\cdot\frac{M_{qm}}{\omega_{q}}.
\end{eqnarray}
The transformed quasi-particles operators have the form
\begin{eqnarray}
\hat{\tilde{\psi}}_{m}=\hat{\psi}_{m}\hat{\chi}_{m},
\end{eqnarray}
where
\begin{eqnarray}
\hat{\chi}_{m}=\exp\left[-\sum_{q}\frac{M_{qm}(\hat{b}_{q}^{\dag}-\hat{b}_{q})}{\omega_{q}}\right].
\end{eqnarray}
The new excitation operator $\hat{\tilde{\psi}}_{m}$ corresponds to a polaron -- quasi-particle coated by phonons. Let us note that the canonical transformation does not change quasi-particle's occupation numbers $\hat{\psi}_{m}^{\dag}\hat{\psi}_{m}=\hat{\tilde{\psi}}_{m}^{\dag}\hat{\tilde{\psi}}_{m}$.
Thus, the Hamiltonian can be written in the following form
\begin{eqnarray}
\hat{H}_{diag}=\sum_{m}(-\Delta_{m})\hat{\tilde{\psi}}_{m}^{\dag}\hat{\tilde{\psi}}_{m}+\hat{H}_{ph},
\end{eqnarray}
where 
\begin{equation}
 \Delta_{m}=4\Delta \cos^{2}(\phi m)
 \label{eq:PolSh}
\end{equation}
is the value of polaron shifts for 2D phonon reservoir with $\Delta=\sum_{q}\frac{M_{q}^{2}}{\omega_{q}}\sim\frac{M^{2}}{s^{2}q_{0}^{2}}\omega_{c}$ and $\omega_c$ is the edge of the phonon spectrum and $s$ is the sound velocity.
Further we will consider $N=2k+1$ and start from the central site of the chain, so $j\rightarrow k+1+j$ where $j=-k\ldots k$, as before. The main formal difference of phonon-assisted hopping from tight-binding model appears due to the quadratic dependence on cosine in the polaron shifts in Eq.~(\ref{eq:PolSh}) absent in the dispersion law, Eq.~(\ref{eq:Em}). 
This model is quite important since it is a clear analogue of classical L\'evy walks due to the presence of real scattering processes. In the initial state only central site $k+1 (j=0)$ is occupied by a quasi-particle.
Using the second quantization representation
\begin{eqnarray}
\psi(0)=c_{0}^{\dag}|g\rangle,
\end{eqnarray}
where $|g\rangle$ is the ground state. The quasi-particle state on the j-site is
\begin{eqnarray}
|j\rangle=c_{j}^{\dag}|g\rangle.
\end{eqnarray}
Thus, the probability to find the quasi-particle on the site $j$ if at the initial time moment $t=0$ it was excited at $j=0$
\begin{eqnarray}
P_{j}(t)=|\langle j|\psi(t)\rangle|^{2}=|\langle g|c_{j}(0)|c_{0}^{\dag}(t)|g\rangle|^{2}.
\end{eqnarray}
As quasi-particles and phonons are decoupled the averaging over quasi-particle and phonon states is independent.
\begin{eqnarray}
&&P(j,t)=\frac{4}{(N+1)^{2}}\sum_{m,m'}\langle\hat{\chi}^{\dag}_{m}(t)\hat{\chi}_{m}(0)\hat{\chi}^{\dag}_{m'}(0)\hat{\chi}_{m'}(t)\rangle\nonumber\\
&&\times \sin(\frac{\pi m}{2})\sin(\frac{\pi m'}{2})\textrm{e}^{i(\Delta_{m}-\Delta_{m'})t}\nonumber\\
&&\times\sin(\frac{\pi m}{2}+\phi mj)\sin(\frac{\pi m'}{2}+\phi m'j). \label{eq:MainPhon}
\end{eqnarray}
Different states $m$ have different polaron shifts $\Delta_{m}$. Hence, the quasi-particle probability distribution $P(j,t)$ acquires oscillating multipliers $\textrm{e}^{i(\Delta_{m}-\Delta_{m'})t}$ arising from averaging over quasi-particle operators (time dependence of quasi-particle Green functions). The influence of phonons is manifested in correlation functions of operators $\langle\hat{\chi}^{\dag}_{m}(t)\hat{\chi}_{m}(0)\hat{\chi}^{\dag}_{m'}(0)\hat{\chi}_{m'}(t)\rangle$. The second order correlation functions of operator $\hat{\chi}_{m}(t)$ can be exactly calculated, as it was done in \cite{Maslova2025}. We consider the situation of zero temperature. In the case of non-zero temperature one can also obtain an exact solution for correlation functions of phonon operators $\hat{\chi}$, but for finite temperature correlation function depends on phonon occupation numbers. The exact expression for correlation function in the case of non-zero temperature can be found in \cite{Maslova2023}. For 2D phonon reservoir this correlation function leads to a power law decay of initial state at difference from 1D and 3D cases \cite{Maslova2023}. As $m,m'=1\ldots N$ one has $\sin(\frac{\pi m}{2})\sin(\frac{\pi m}{2}+\phi mj)=\sin^{2}(\frac{\pi m}{2})\cos(\phi mj)$ putting Eq.~(\ref{eq:MainPhon}) into the class discussed in Sec.~\ref{sec:Hetero}. Taking into account that $\sum_{q}\frac{M_{q}^{2}}{\omega_{q}^{2}}(1-\textrm{e}^{-i\omega_{q}t})=\alpha \ln(\omega_{c}t)$ with $\alpha=\frac{M^{2}}{q_{0}^{2}s^{2}}$ and $\Delta\sim\alpha\omega_{c}$ one can obtain
\begin{eqnarray}
&&P(j,t)=\frac{4}{(N+1)^{2}}\sum_{mm'=1}^{N} \cos(\phi mj)\cos(\phi m'j) \nonumber\\
&& \qquad \times \exp \big\{i\alpha(\omega_{c}t)[\cos(2\phi m)-\cos(2\phi m')] \nonumber \\
&& \qquad \qquad  -\alpha \ln (\omega_{c}t) [ \cos(\phi m)-\cos(\phi m')]^{2} \big\} \label{eq:PhonSum}
\end{eqnarray}
The sum can be approximated by an integral which can be evaluated for $\omega_{c}t\gg1$ using the saddle-point approximation. Introducing $z=\omega_{c}t$, one gets
\begin{eqnarray}
P(j,t) \simeq &&  \frac{1}{2\alpha z} \left[1-\left(\frac{j}{2\alpha z}\right)^{2}\right]^{-1/2} \nonumber \\
&&  \times \left\{1+\frac{\cos(\pi j)}{z^{2\alpha}}z^{-2\alpha\left[1-(\frac{j}{2\alpha z})^{2}\right]^{1/2}}\right\}
\label{eqn:qppdf}
\end{eqnarray}
for $j<2\alpha z$. For $j>2\alpha z$ $P_{j}=0$. 
The quasi-particle probability density function has the U-like shape with a well-pronounced front, see Fig.~\ref{fig:fig_phonons}.  

The spread of the distribution can be obtained using Eq.~(\ref{eq:Width2}) and is given by 
\begin{equation}
 W^2(t) = \alpha [2 \omega_c t + \ln(\omega_c t)],
\end{equation}
and is ballistic with a logarithmic correction. 

The PDF for the 2D-phonon-driven quasi-particles resembles the one obtained for tight-binding model, and is similar to the one of classical L\'evy walks model in its ballistic regime. The reason for such similarity is that interaction with 2D phonon reservoir uniform for all the sites leads to appearance of strong ballistic component in quasi-particle transport due to limited velocity of quasi-particles spreading, and scattering. For 1D and 3D phonon reservoir the shape of PDF changes byt the spread remains ballistic.

Quasi-particle operators $\hat{c}_{j}/\hat{c}_{j}^{\dag}$ do not allow to properly describe elementary excitations. Genuine elementary excitations are polarons $\hat{\tilde{\psi}}_{m}$ -- the quasi-particles (say, electrons) coated by phonons. Polarons dispersion law is given by $\Delta_{m}$. Setting a quasi-particle on the site $l$ disturbs the phonon subsystem and leads to the appearance of a scattering potential $\hat{V}$ 
\begin{eqnarray}
\hat{V}=\hat{c}_{l}^{\dag}\hat{c}_{l}=\sum_{mm'}\textrm{e}^{i\phi(m-m')l}\hat{\psi}_{m}^{\dag}\hat{\psi}_{m'}\nonumber\\=
\sum_{mm'}\hat{\tilde{\psi}}_{m}^{\dag}\hat{\tilde{\psi}}_{m'}\hat{\chi}_{m}^{\dag}\hat{\chi}_{m'}\textrm{e}^{i\phi(m-m')l}=\sum_{mm'}\hat{V}_{mm'},
\end{eqnarray}
where
\begin{eqnarray}
\hat{\chi}_{m}^{\dag}(t)\hat{\chi}_{m'}(t)=\textrm{exp}[\sum_{q}\frac{M_{q}}{\omega_{q}}(\hat{b}_{q}^{\dag}(t)-\hat{b}_{q}(t))\nonumber\\\cdot(\textrm{cos}(\phi m)-\textrm{cos}(\phi m'))].\nonumber\\
\end{eqnarray}

Thus, after the quench the Hamiltonian has the form
\begin{eqnarray}
\hat{H}_{diag}=\sum_{m}(-\Delta_{m})\hat{\tilde{\psi}}_{m}^{\dag}\hat{\tilde{\psi}}_{m}+\hat{H}_{ph}+\hat{V}.
\end{eqnarray}
The appearance of potential $\sum_{mm'}\hat{V}_{mm'}$ causes polaron scattering. Scattering of polaron is more effective from state $m$ to state $m'=N+1-m$ (it is the case of backscattering). So, strong ballistic component with backscattering on the random potential appears in quasi-particle transport. It is a direct quantum analogue of L\'evy walks. Since back scattering of polarons is present in the considered model. Probability distribution function $P(j,t)$ differs from anomalous diffusive case with waiting time probability function proportional to $(\omega_{c}t)^{-1-\alpha}$ \cite{Maslova2025}. Superposition of quantum oscillations with different frequencies $(\Delta_{m}-\Delta_{m'})$ leads to the additional power law damping and drastically changes the shape of $P(j,t)$. Results within the diffusive approximation can be obtained only if each quasi-particle state acquire its own independent random phase $\delta_{m}$ for each state $m$. To conclude, the time dependent PDF for phonon assisted spreading of quasi-particles is directly the result of two net effects: the presence of quantum coherence due to the interaction with the same reservoir and the limited velocity of quasi-particles spreading.

\begin{figure}[h]
\centering
\includegraphics[width=\linewidth]{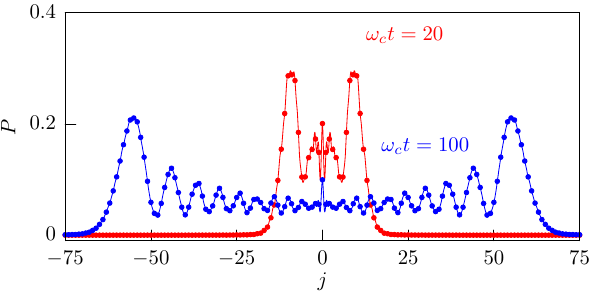}%
\caption{Quasi-particle probability density function for initially excited quasi particles at site $j=0$ and for different values of parameter $\omega_{c}t$ and $\alpha=0.3$. The uniform phase $\delta$ is distributes as $P(\delta)=1/2\pi$. } 
\label{fig:fig_phonons_phase}
\end{figure}

Let us show that for phonon-assisted hopping the uniform random phase $\delta$ (equal for all the sites) with uniform probability density $P(\delta)=1/2\pi$ does not change anything. 
Diagonalization of Hamiltonian for quasi-particles can be done in the following way:
\begin{eqnarray}
\hat{c}_{j}=\sqrt{\frac{2}{N+1}}\sum_{m}\sin(m\phi j+\delta j)\hat{\psi}_{m},
\end{eqnarray}
so that the Hamiltonian has the form
\begin{eqnarray}
\hat{H}=\sum_{m}2\hat{\eta}\cos(\phi m+\delta)\hat{\psi}_{m}^{\dag}\hat{\psi}_{m}+\hat{H}_{ph}
\end{eqnarray}
with 
\begin{eqnarray}
\hat{\eta}=\sum_{q}M_{q}(\hat{b}_{q}+\hat{b}_{q}^{\dag}).
\end{eqnarray}
Using the same decoupling procedure as for the case $\delta=0$ considered above, we obtain 
\begin{eqnarray}
P_{j}(t)&&=\frac{2}{\pi(N+1)^{2}}\int_{-\pi}^{\pi}d\delta\sum_{mm'}\cos(\phi m j)\cos(\phi m' j)\nonumber\\
&&\times\textrm{exp}[i\alpha\omega_{c}t(\cos(2\phi m+2\delta)-\cos(2\phi m'+2\delta))\nonumber\\&&-\alpha \ln(\omega_{c}t)(\cos(\phi m+\delta)-\cos(\phi m'+\delta))^{2}].
\label{eqn:26}
\end{eqnarray}
For the case $\delta=0$ discussed before one omits the outer integral over $\delta$, and obtains an alternative representation of Eq.~(\ref{eq:PhonSum}).
Changing $\sum_{mm'}\rightarrow\int_{0}^{\pi}\int_{0}^{\pi}dxdy$ and performing substitution $\frac{u-v}{2}=x$, $\frac{u+v}{2}=y$ one can get
\begin{eqnarray}
P_{j}(t)&&=\frac{1}{2\pi}\int_{-\pi}^{\pi}d\delta\int_{-\pi}^{\pi}dv\int_{0}^{2\pi}du[\cos(vj)+\cos(uj)]\nonumber\\&&
\times\textrm{exp} \big[i2\alpha\omega_{c}t\cdot \sin(u+2\delta)\sin(v)\nonumber\\&&-4\alpha \ln(\omega_{c}t) \cdot \sin^{2}(u/2+\delta)\sin^{2}(v/2) \big].
\end{eqnarray}
For the case $\delta=0$, one omits the outer integral over $\delta$, and puts $\delta = 0$ in the arguments of the trigonometric functions in the second and the third lines of the last expression. One obtains 
\begin{eqnarray}
P_{j}(t)&&=2\int_{-\pi}^{\pi}dv\int_{0}^{2\pi}du \cos(vj)
\textrm{exp}[i2\alpha\omega_{c}t\cdot \sin(u)\sin(v)\nonumber\\&&-4\alpha \ln(\omega_{c}t) \cdot \sin^{2}(u/2)\sin^{2}(v/2)].
\end{eqnarray}
For $\delta\neq0$ after changing the variable to $\tilde{u}/2=u/2+\delta$, one gets
\begin{eqnarray}
P_{j}(t)&&=\frac{1}{2\pi}\int_{-\pi}^{\pi}d\delta\int_{-\pi}^{\pi}dv\int_{0}^{2\pi}d\tilde{u}[\cos(vj)+\cos((\tilde{u}-2\delta)j)]\nonumber\\
&& \times\exp \big[i2\alpha\omega_{c}t\cdot \sin(\tilde{u})\sin(v) \\
&& \;\; -4\alpha \ln(\omega_{c}t) \cdot \sin^{2}(\tilde{u}/2)\sin^{2}(v/2) \big]. \nonumber
\end{eqnarray}
The PDF includes two terms
\begin{eqnarray}
P_{j}(t)=K_{1}(t)+K_{2}(t).
\end{eqnarray}
The first one  $K_{1}(t)$ has the form 
\begin{eqnarray}
K_{1}(t)&&=\frac{1}{2\pi}\int_{-\pi}^{\pi}d\delta\int_{-\pi}^{\pi}dv\int_{-2\pi}^{4\pi}d\tilde{u} \cos(vj)\nonumber\\
&&\times\textrm{exp} \big[i2\alpha\omega_{c}t\cdot \sin(\tilde{u})\sin(v) \\
&& \qquad -4\alpha \ln(\omega_{c}t) \cdot \sin^{2}(\tilde{u/2})\sin^{2}(v/2)\big] \sim P_{j}(t)|_{\delta=0}. \nonumber
\end{eqnarray}
The second term $K_{2}(t)$ can be expressed as
\begin{eqnarray}
K_{2}(t)=K_{2}^{'}(t)+K_{2}^{''}(t)
\end{eqnarray}
with
\begin{eqnarray}
K_{2}^{'}(t)&&=\frac{1}{2\pi}\int_{-\pi}^{\pi}d\delta\cos(2\delta j)\int_{-\pi}^{\pi}dv\int_{-2\pi}^{4\pi}d\tilde{u} \cos(\tilde{u}j)\nonumber\\
&&\times\textrm{exp} \big[i2\alpha\omega_{c}t\cdot \sin(\tilde{u})\sin(v) \nonumber \\
&& \qquad -4\alpha \ln(\omega_{c}t) \cdot \sin^{2}(\tilde{u}/2)\sin^{2}(v/2)\big] = 0\nonumber
\end{eqnarray}
as integrals over $v$ and $\tilde{u}$ does not depend on $\delta$.
Analogously,
\begin{eqnarray}
K_{2}^{''}(t)&&=\frac{1}{2\pi}\int_{-\pi}^{\pi}d\delta \sin(2\delta j)\int_{-\pi}^{\pi}dv\int_{-2\pi}^{4\pi}d\tilde{u} \sin(\tilde{u}j)\nonumber\\
&&\times\textrm{exp} \big[i2\alpha\omega_{c}t\cdot \sin(\tilde{u})\sin(v) \nonumber \\
&& \qquad -4\alpha \ln(\omega_{c}t) \cdot \sin^{2}(\tilde{u}/2)\sin^{2}(v/2)\big] = 0\nonumber
\end{eqnarray}
The same random phase for all the sites does not change anything for uniform phonon bath. The direct evaluation of $P_{j}(t)$ as given by Eq.~(\ref{eqn:26}) is shown in Fig.~\ref{fig:fig_phonons_phase}.

We would like to estimate the typical times $t_{j}$ of quasi-particles spreading from the initially excited site $j=0$ at $t=0$ to the site $j$. An expression for $t_{j}$ reads $t_{j}=\frac{j}{\omega_{c}\alpha}$. For $\alpha=0.1$ \cite{Maslova2025} and $\omega_{c}\simeq10^{13}$ $sec^{-1}$ \cite{Semina2018} one can get $t_{j}\sim10^{-12}j$ sec. For $j\simeq10\div100$ one can get $t_{j}\sim10\div100$ picoseconds, which is a typical time scale for exciton transport in transition metal dichalcogenides mono- and bi-layers \cite{Chernikov2023}. 

\section{Conclusions \label{sec:Concl}}
\label{Conclusions}
We analyzed the spreading of locally excited quasiparticles in several quantum systems: i) quasiparticle spreading along a chain in the simple tight-binding model; ii) quasiparticle transport in a heterogeneous ensemble of chains with a Gaussian distribution of hopping amplitudes; iii) quasiparticle spreading in a tight-binding model with classical time-dependent random hopping amplitude and an arbitrary correlation function; iv) phonon-assisted quasiparticle spreading along a chain coupled to a uniform phonon reservoir.

For the simple tight-binding model, the probability distribution function (PDF) of a locally excited quasiparticle is U-shaped and resembles that of classical L\'evy walks. The spread of the PDF is ballistic, and this behavior arises from the strictly limited velocity of quasiparticle propagation. In the heterogeneous ensemble of chains with a Gaussian distribution of hopping amplitudes, the time-dependent PDF transforms from bimodal to monomodal via an intermediate trimodal shape as the parameters of the Gaussian distribution are varied. The spread of the PDF remains ballistic throughout. For the model with random time-dependent classical hopping amplitude, the PDF of quasiparticle spreading exhibits universal asymptotics with logarithmic corrections and resembles the diffusing diffusivity model. The spread of the PDF is diffusive. In the case of phonon-assisted hopping of quasiparticles coupled to a common phonon reservoir, the PDF also resembles that of classical L\'evy walks. It is U-shaped due to the presence of a strong ballistic component in the spreading dynamics. This behavior emerges from quantum coherence induced by the interaction with the same reservoir, combined with the fact that the absolute value of the quasiparticle velocity is bounded from above. Unlike the simple tight-binding model, phonon-assisted hopping along the chain represents a evident quantum analog of classical L\'evy walks, as it involves real scattering processes.

Furthermore, we revealed that the microscopic origin of heavy-tailed waiting time distributions can be attributed to the interaction with the same two-dimensional phonon reservoir.

\section{Acknowledgements}

N.S.M. and V.N.M. thank for support Russian Science Foundation grant No. 24-12-00020.% We also thank the Foundation for the Advancement of Theoretical Physics and Mathematics ``BASIS''.


\begin{thebibliography}{99}

\bibitem{Mueller2018} T. Mueller, and E. Malic, Exciton physics and device application of two-dimensional transition metal dichalcogenide semiconductors, npj 2D Mater. Appl.  {\bf 2}, 29 (2018).
\bibitem{Mishra2020} P. Mishra, D. Singh, Y. Sonvane, and R. Ahuja, Excitonic effects in the optoelectronic properties of graphene-like BC monolayer, Opt. Mater. {\bf 110}, 110476 (2020).
\bibitem{Fu2023} J. Fu, S. Ramesh, J. W. M. Lim, and T. C. Sum, Carriers, Quasi-particles, and Collective Excitations in Halide Perovskites, Chem. Rev. {\bf 123}, 8154 (2023).
\bibitem{Pokryshkin2023} N.S. Pokryshkin, V.N. Mantsevich, and V. Y. Timoshenko, Carriers, Anti-Stokes Photoluminescence in Halide Perovskite Nanocrystals: From Understanding the Mechanism towards
Application in Fully Solid-State Optical Cooling, Nanomaterials {\bf 13}, 1833 (2023).
\bibitem{Causin2022} R. Perea-Causin, D. Erkensten, J.M. Fitzgerald, J.J.P. Thompson, R. Rosati, S. Brem, and E. Malic, Carriers, Exciton optics, dynamics, and transport in atomically thin semiconductors, APL Mater. {\bf 10}, 100701 (2022).
\bibitem{Sheehan2024} T.J. Sheehan, S. Saris, and W.A. Tisdale, Carriers, Exciton Transport in Perovskite Materials, Adv. Mater. {\bf 37}, 2415757 (2024).
\bibitem{Zaburdaev2015} V. Zaburdaev, S. Denisov, and J. Klafter, L\'evy walks, Rev. Mod. Phys. {\bf 87}, 483 (2015).
\bibitem{Klafter1987} J. Klafter, A. Blumen, and M. F. Shlesinger, Stochastic pathway to anomalous diffusion, Phys. Rev. A {\bf 35}, 3081 (1987).
\bibitem{Lamperti1958} J. Lamperti, An occupation time theorem for a class of stochastic processes, Trans. Amer. Math. Soc. {\bf 88}, 380 (1958).
\bibitem{Nirmal1996} M. Nirmal, B. O. Dabbousi, M. G. Bawendi, J. J. Macklin, J. K. Trautman, T. D. Harris, and L. E. Brus, Fluorescence intermittency in single cadmium selenide nanocrystals, Nature {\bf 383}, 802 (1996).
\bibitem{Empedocles1996} S. A. Empedocles, D. J. Norris, and M. G. Bawendi, Photoluminescence spectroscopy of single CdSe nanocrystallite quantum dots, Phys. Rev. Lett. {\bf 77}, 3873 (1996).
\bibitem{Bischof2014} T. S. Bischof, R. E. Correa, D. Rosenberg, E. A. Dauler, and M. G. Bawendi, Measurement of emission lifetime dynamics and biexciton emission quantum yield of individual InAs colloidal nanocrystals, Nano Lett. {\bf 14}, 6787 (2014).
\bibitem{Protasenko2005} V. V. Protasenko, K. L. Hull, and M. Kuno, Disorder-induced optical heterogeneity in single CdSe nanowires, Adv. Mater. {\bf 17}, 2942 (2005).



\bibitem{Margolin2006} G. Margolin, V. Protasenko, M. Kuno, and E. Barkai, Photon counting statistics for blinking CdSe-ZnS quantum dots: a L\'evy walk process, J Phys. Chem. B {\bf 110}, 19053 (2006).
\bibitem{Kutner1997} R. Kutner, and P. Maass, Random walk on a linear chain with a quenched distribution of jump lengths, Phys. Rev. E {\bf 55}, 71 (1997).
\bibitem{Barkai2020} E. Barkai, and S. Burov, Packets of diffusing particles exhibit universal exponential tails, Phys. Rev. Lett. {\bf 124}, 060603 (2020).





\bibitem{Novoselov2004} K. S. Novoselov, A. K. Geim, S. V. Morozov, D. Jiang, Y. Zhang, S. V. Dubonos, I. V. Grigorieva, and A. A. Firsov, Electric Field Effect in Atomically Thin Carbon Films  Science {\bf 306}, 666 (2004).
\bibitem{Li2024} Y. Li, J. Cao, G. Chen, L. He, X. Du, J. Xie, Y. Wang, and W. Hu, Scalable Production of Highly Conductive 2D NbSe2 Monolayers with Superior Electromagnetic Interference Shielding Performance ACS Appl. Mater. Interfaces {\bf 16}(5), 6250 (2024).
\bibitem{Manzeli2017} S. Manzeli, D. Ovchinnikov, D. Pasquier, O. V. Yazyev, and A. Kis, 2D transition metal dichalcogenides Nat. Rev. Mater.{\bf 2}(5), 17033 (2017).
\bibitem{Molaei2021} M. J. Molaei, M. Younas, and M. Rezakazemi, A Comprehensive Review on Recent Advances in Two-Dimensional (2D) Hexagonal Boron Nitride ACS Appl. Electron. Mater.{\bf 3}(12), 5165 (2021).
\bibitem{Zhang2021} S. Zhang, R. Xu, N. Luo, and X. Zou, Scalable Two-dimensional magnetic materials: structures, properties and external controls Nanoscale {\bf 13}, 1398 (2021).
\bibitem{Shanmugam2022} V. Shanmugam, R. A. Mensah, K. Babu, S. Gawusu, A. Chanda, Y. Tu, R. E. Neisiany, M. Försth, G. Sas, O. Das, A Review of the Synthesis, Properties, and Applications of 2D Materials. Part. Part. Syst. Charact. {\bf 39}, 2200031 (2022).
\bibitem{Frenkel1931} J. Frenkel, On the transformation of light into heat in solids. Phys. Rev. {\bf 37}(1), 17 (1931).
\bibitem{Gross1952} E. F. Gross, N. A. Karrijew, Light absorption by cuprous oxide crystal in infrared and visible part of the spectrum. Dokl. Akad. Nauk SSSR {\bf 84}, 471 (1952).
\bibitem{Xiao2017} J. Xiao, M. Zhao, Y. Wang, and X. Zhang, Excitons in atomically thin 2D semiconductors and their applications. Nanophotonics {\bf 6}, 1309 (2017).
\bibitem{Wang2018} G. Wang, A. Chernikov, M. M. Glazov, T. F. Heinz, X. Marie, T. Amand, and B. Urbaszek, Colloquium: excitons in atomically thin transition metal dichalcogenides. Rev. Modern Phys. {\bf 90}(2), 021001 (2018).
\bibitem{Kumar2014} N. Kumar, Q. Cui, F. Ceballos, D. He, Y. Wang, H. Zhao, Exciton diffusion in monolayer and bulk MoSe2. Nanoscale {\bf 6}, 4915 (2014).
\bibitem{Kurilovich2022} A.A. Kurilovich, V. N. Mantsevich, Y. Mardoukhi, K. J. Stevenson, A. V. Chechkin, and V. V.  Palyulin, Non-Markovian diffusion of excitons in layered perovskites and transition metal dichalcogenides. Phys. Chem. Chem. Phys. {\bf 24}(22), 13941 (2022).
\bibitem{Kurilovich2024} A.A. Kurilovich, V. N. Mantsevich, A. V. Chechkin, and V. V.  Palyulin, Negative diffusion of excitons in quasi-twodimensional
systems. Phys. Chem. Chem. Phys. {\bf 26}, 922 (2024).
\bibitem{Wietek2024} E. Wietek, M. Florian, J. Goser, T. Taniguchi, K. Watanabe, A. Hogele, M. M. Glazov, A. Steinhoff, and A. Chernikov, Nonlinear and negative effective diffusivity of interlayer excitons in Moire-free heterobilayers, Phys. Rev. Lett. {\bf 132}, 016202 (2024).
\bibitem{Fogler2014} M. M. Fogler, L. V. Butov, and K. S. Novoselov, Hightemperature superfluidity with indirect excitons in van der Waals heterostructures, Nat. Commun. {\bf 5}, 4555 (2014).
\bibitem{Aguila2023} A. G. del Aguila, Y. R. Wong, I. Wadgaonkar, A. Fieramosca, X. Liu, K. Vaklinova, S. Dal Forno, T. Thu Ha Do, H. Y. Wei, K. Watanabe, T. Taniguchi, K. S. Novoselov, M. Koperski, M. Battiato, and Q. Xiong, Ultrafast exciton fluid flow in an atomically thin MoS2 semiconductor, Nat. Nanotechnol.{\bf 18}, 1012 (2023).
\bibitem{Mantsevich2025} V.N. Mantsevich, and M.M. Glazov, Viscous hydrodynamics of excitons in van derWaals heterostructures, Phys. Rev. B {\bf 110}, 165305 (2024)
\bibitem{Dubin2005} F. Dubin, R. Melet, T. Barisien, R. Grousson, L. Legrand, M. Schott, and V. Voliotis, Macroscopic coherence of a single exciton state in an organic quantum wire, Nature Phys. {\bf 2}, 32 (2006).
\bibitem{Maslova2025} N. S. Maslova, V. N. Mantsevich, and P. I. Arseyev, Exciton transport in atomically flat heterostructures: The appearance of negative diffusivity, Phys. Rev. E {\bf 112}, 014102 (2025).
\bibitem{Maslova2023} N. S. Maslova, P. I. Arseyev, I. M. Sokolov, and V. N. Mantsevich, Noise induced dynamics of two-qubit entangled Bell’s states, J. Phys. Chem. Solid {\bf 183}, 111638 (2023).
\bibitem{Semina2018} S. Shree, M. Semina, C. Robert, B. Han, T. Amand, A. Balocchi, M. Manca, E. Courtade, X. Marie, T. Taniguchi, K. Watanabe, M. M. Glazov, and B. Urbaszek, Observation of
exciton-phonon coupling in $MoSe_2$ monolayers, Phys. Rev. B {\bf 98}, 035302 (2018).
\bibitem{Chernikov2023} A. Chernikov, and M. M. Glazov, Exciton diffusion in 2D van der Waals semiconductors, Semiconductors and Semimetals, {\bf 112}, 69 (2023).
\bibitem{Linares} J. Linares, B. Atenas, and S. Curilef, Fluctuations, Complexity and Statistical Measures for Particles in
a Tight-Binding Lattice, Symmetry \textbf{18}, 493 (2026).
\bibitem{Cuevas} F.A. Cuevas, S. Curilef, and A.R. Plastino, Spread of highly localized wave-packet in the tight-binding
lattice: Entropic and information-theoretical characterization, Annals of Physics \textbf{326}, 2834 (2011).
\bibitem{Penson} U. Busch and K.A. Penson, Tight-binding electrons on open chains: Density distribution and correlations, Phys. rev. B \textbf{36}, 9271 - 9274 (1987).
\bibitem{AbraSteg} M. Abramowitz and I.A. Stegun, \textit{Handbook of Mathematical Functions with Formula, Graphs and Mathematical Tables}, National Bureau of Standards, Dover (1972). 
\bibitem{BPM2} A.P. Prudnikov, Yu.A. Brychkov and O.I. Marichev, Integrals and Series: Special functions, Gordon and Breach, Amsterdam, 1998
\bibitem{Chubyns} M.V. Chubynsky and G.W. Slater, Diffusing Diffusivity: A Model for Anomalous, yet Brownian, Diffusion, Phys. Rev. Lett. \textbf{113}, 098302 (2014).
\bibitem{ChechSok} A.V. Chechkin, F. Seno, R. Metzler, and I.M. Sokolov, Brownian yet non-Gaussian diffusion: from superstatistics to subordination of diffusing diffusivities, Physical Review X \textbf{7}, 021002 (2017).
\bibitem{Zwanzig} R. Zwanzig, \textit{Nonequilibrium Statistical Mechanics}, Oxford University Press, Oxford, (2001).
\bibitem{Wang} B. Wang, S.M. Anthony, S.C. Bae, and S. Granick, Anomalous Yet Brownian, Proc. Natl. Acad. Sci. U.S.A. \textbf{106}, 15160 (2009).



\end{thebibliography}
\end{document}